\documentclass[final]{siamltex}
\usepackage{epsfig, amsmath, amssymb}%,fullpage}
\usepackage[colorlinks]{hyperref}
\usepackage[lined,boxed]{algorithm2e} %{algorithm2e}
\usepackage{multirow}
\newcommand{\mtp}{minimum $p$-sum problem}

\newcommand{\scotch}{\textsc{Scotch}}

\title{\Large Relaxation-based coarsening and multiscale graph organization}
\author{
Dorit Ron\thanks{Faculty of Mathematics and Computer Science, The Weizmann Institute of Science, \href{mailto:dorit.ron@weizmann.ac.il}{\tt dorit.ron@weizmann.ac.il}}
\and Ilya Safro\thanks{Mathematics and Computer Science Division, Argonne National Laboratory, \href{mailto:safro@mcs.anl.gov}{\tt safro@mcs.anl.gov}}
\and Achi Brandt\thanks{Faculty of Mathematics and Computer Science, The Weizmann Institute of Science, \href{mailto:achi.brandt@weizmann.ac.il}{\tt achi.brandt@weizmann.ac.il}}
}

\begin{document}

\maketitle
\begin{abstract}
In this paper we generalize and improve the multiscale organization of graphs by introducing a new measure that quantifies the ``closeness'' between two nodes. The calculation of the measure is linear in the number of edges in the graph and involves just a small number of relaxation sweeps. A similar notion of distance is then calculated and used at each coarser level. We demonstrate the use of this measure in multiscale methods for several important combinatorial optimization problems and discuss the multiscale graph organization.
\end{abstract}

\section{Introduction}

A general approach for solving many large-scale graph problems, as
well as most other classes of large-scale computational science problems, is
through multilevel (multiscale, multiresolution, etc.)
algorithms. This approach generally involves {\it coarsening}
the problem, producing from it a sequence of progressively coarser
levels (smaller, hence simpler, related problems), then
recursively using the (approximate) solution of each coarse
problem to provide an initial approximation to the solution at the
next-finer level. At each level, this initial approximation is first
improved by what we generally call ``local processing" (LP). This
is an inexpensive sequence of short steps, each involving only a few
unknowns, together covering all unknowns of that level several
times over. The usual examples of LP are few sweeps of classical
(e.g., Gauss-Seidel or Jacobi) relaxation in the case of solving a
system of equations, or a few Monte Carlo passes in
statistical-physics simulations. Following the LP, the resulting approximation
may be further improved by one or several cycles, each
using again a coarser-level approximation followed by LP, applying  them at each time to
the {\it residual} problem (the problem of calculating the {\it
error} in the current approximation). See, for example, references
\cite{Brandt:1977:MLAa,bamg1,bmr82,bmr84,brandt-renormalization,vlsicad,rs87,safro2005}.

At each level of coarsening one needs to define the set of  coarse
unknown variables and the equations (or the stochastic relations)
that they should satisfy (or the energy that they should
minimize). Each coarse unknown is defined in terms of the
next-finer-level unknowns ({\it defined}, not {\it calculated}: they  are all unknowns until the coarse level is approximately solved and the fine level is interpolated from
that solution). The following are examples:
\begin{itemize}
    \item The set of coarse unknowns can simply represent a chosen
    subset of the fine-level set.
    \item If the fine-level variables are real numbers or vectors,
    each coarse variable can represent a weighted average of
    several of them.
    \item If the fine-level variables are Ising spins (having only
    values of $+1$ or $-1$), each coarse variable can again be an
    Ising spin, representing the {\it sign} of the sum of  several
    fine spins.
    \item A coarse variable can be defined from several fine
    variables by a stochastic process (\cite{blote96}, for example).
    \item In the case of graph problems, each node of the coarse
    graph can represent an {\it aggregate} of several fine-level
    nodes or a {\it weighted aggregate} of such nodes, that is,
    allowing each fine-level node to be split between several
    aggregates.
\end{itemize}

The choice of an adequate local processing at a fine level
and the choice of an adequate set of variables at the next-coarser level are
strongly coupled. The general guiding rule \cite{SU} is
that this pair of choices is good if (and to the extent that) a
fine-level solution can always be recovered from the corresponding
set of coarse variables by a short iterative use of a suitably
modified version of the LP. That version is called {\it compatible LP} (CLP).
Examples are compatible Monte Carlo (CMC), introduced in
\cite{brandt-renormalization}, and compatible relaxation (CR),
introduced in \cite{crs}.

The CLP, needed in several important upscaling procedures
(such as the selection of the coarse variables, the acceleration of the fine-level simulations,
and the processing of fine-level windows within coarse simulations; see
\cite{SU}) can also be used for performing the interpolation
from the coarse solution to obtain the first approximation at the
fine level. When possible, however, the construction of a more
explicit interpolation is desired in order to apply it for the direct formulation of
equations (or an energy functional) that should govern the
coarse level, as in Galerkin coarsening.

In the process of defining the set of coarse variables and
in constructing an explicit interpolation, it is important to know
how ``close"  two given fine-level variables are to each other at
the stage of switching to the coarse level. We need to know, in other words, to what extent
the value after the LP of one variable implies the value of the other.
If they are sufficiently close, they can, for example, be
aggregated to form a coarse variable.

The central issue addressed in the present article is how to
measure this ``closeness" between two variables in a system of
equations or between two nodes in a given graph. (We consider the
latter to be a special case of the former, by associating the
graph with the system formed by its Laplacian.) More generally, we
want to define the distance of one variable $x_i$ from a small subset
$S$ of {\it several} variables, in order to measure how well $x_i$
can be interpolated from $S$ following the LP.

In classical Algebraic Multigrid (AMG), aimed at solving the
linear system
\begin{equation}\label{AMG1}
Ax = b~~~~~ {\text or}~~~~~ \sum_{j=1}^n a_{ij} x_j = b_i ~~,~~
(i= 1, ... ,n)~~,
\end{equation}
the closeness of two unknowns $x_i$ and $x_j$ is measured simply
by the relative size of their coupling $a_{ij}$, for example, by the quantity
\begin{equation}\label{AMG2}
|a_{ij}| / {\text {max}}(\sum_k |a_{ik}| , \sum_k |a_{kj}|)
\end{equation}
(or similarly by the relative size of their coupling in some {\it
power} of $A$). Although this definition has worked well for
the coarsening procedures of discretized scalar elliptic
differential equations, it is not really effective, and sometimes
meaningless, for systems lacking sufficient diagonal
dominance (including many discretized {\it nonscalar} elliptic
systems). Moreover, even for systems with a fully dominant diagonal
(such as the Laplacian of a graph), the classical AMG definition
may result in wrong coarsening, for example, in graphs
with nonlocal edges (see example in Sec. \ref{algorithm}).

Instead, we propose to define the ``closeness" between two
variables exactly, by measuring how well their values are
correlated at the coarsening stage, namely, following the LP
relaxation sweeps. Since the coarse level is actually applied to
the {\it residual} system, the two variables will be considered
close if their {\it errors} have nearly the same ratio in all
relaxed vectors. We will thus create a sequence of $K$ {\it
normalized relaxed error vectors} $x^{(1)}, ... ,x^{(K)}$, each
obtained by relaxing the homogeneous system $Ax = 0$ from some
(e.g., random) start and then normalizing the result. We will then
define the {\it algebraic distance} (reciprocal of ``closeness")
between any two variables $x_i$ and $x_j$ as
\begin{equation}\label{intro3}
{\text {min}}_{\eta} ~\bigg(~\sum_{k=1}^K |\eta x_i^{(k)} - {|\eta^{-1}|
x_j^{(k)}|^p~\bigg)^{1/p}}~~~~,
%~~~~(p=2 ~~{\text {or}} ~~\infty)~.
\end{equation}
where $p \geq 2$ in order to attach larger weights to larger differences (using usually either $p=2$ or the maximum norm ($p \rightarrow \infty$)). This use of $\eta$ gives a symmetric measure of how well
$x_i$ can be interpolated from $x_j$ or vice versa.
For the graph Laplacian (and other zero-sum $A$) this can
be simplified to a distance defined as
\begin{equation}\label{intro4}
\bigg(~\sum_{k=1}^K (x_i^{(k)}-x_j^{(k)})^2~\bigg)^{1/2} ~~~~~{\text {or}}
~~~~~\max_{k=1}^K |x^{(k)}_i - x^{(k)}_j|~.
\end{equation}

More generally, the distance of a node $x_i$ from a subset $S$ of
several nodes can similarly be defined as the deviation of the
best-fitted interpolation from $S$ to $x_i$, where the deviation
is the $L_2$ norm of the vector of $K$ errors obtained upon
applying the interpolation to our $K$ normalized relaxed error
vectors, and the best-fitted interpolation is the one having the
minimal deviation. (This least-square interpolation is the one
introduced in bootstrap AMG (BAMG) \cite{amg} for the coarse-to-fine explicit
interpolation.)

An essential aspect of the ``algebraic distance" defined here is
that it is a crude {\it local distance}. It measures
meaningful closeness only between neighboring nodes; the closer
they are the less fuzzy is their measured distance. For nodes that
should not be considered as neighbors, their algebraic distance
just detects the fact that they are far apart; its exact value carries
no further meaning. The important point is that this crude local
definition of distance is fast to calculate and is all
that is required for the coarsening purposes. A similar notion of
distance is then similarly calculated at each coarser level.

Indeed, we argue that meaningful distances in a general graph
should, in principle, be {\it defined} (not just {\it calculated})
only in such a multiscale fashion. This essential viewpoint, and
relations to diffusion distances and spectral clustering are
discussed in Section \ref{multiscale}. In particular, we
advocate the replacement of spectral methods by AMG-like
multilevel algorithms, which are both faster and more tunable to
define better solutions to many fuzzy graph problems
(see, for example, \cite{safro2004,safro2005}).

The paper is organized as follows. The graph problems we use to
demonstrate our approach are introduced in Sec \ref{prob-def}. The
calculation of the ``algebraic distance" and its use within the
multiscale algorithm is described in Section \ref{algorithm}. Results of tests
are summarized in Section \ref{results}. Finally, the
relations of our approach to diffusion distances and spectral
clustering are discussed in Section \ref{multiscale}.

\section{Notation and problem definitions}\label{prob-def}
\par Given a weighted graph $G=(V,~E)$, where $V=\{1,2,...,n\}$
is the set of nodes (vertices) and $E$ is the set of edges.
Denote by $w_{ij}$ the non-negative weight (coupling)
of the undirected edge $ij$ between nodes $i$ and $j$; if
$ij\notin E$, then $w_{ij}=0$. We consider as our examples the following two optimization problems.
\subsection{Linear ordering}
Let $\pi$ be a bijection
\[
\pi : ~ V ~ \longrightarrow ~ \{1,2,...,n\}~~~.
\]
The purpose of linear ordering problems is to minimize some functional over all possible permutations $\pi$. The following functional should be minimized for the {\mtp}\footnote{We use this definition for simplicity. The usual definition of the functional is $\sigma_p(G,\pi)=(\sum_{ij}w_{ij}|\pi(i)-\pi(j)|^p)^{1/p}$, which yields the same minimization problem.} (M$p$SP):
\begin{equation}\label{minpsumdef}
\sigma_p(G,\pi)=\sum_{ij}w_{ij}|\pi(i)-\pi(j)|^p~~~.
\end{equation}
 In the generalized form of the problem that emerges during the multilevel solver, each vertex $i$ is assigned with a $volume$ (or $length$), denoted $v_i$. Given the vector of all volumes, $v$, the task now is to minimize the cost
 \[
 \sigma_p(G, \pi, v)\overset{def}{=} \sigma_p(G,x)=\sum_{ij}w_{ij}|x_i-x_j|^p~~~,
 \]
where \mbox{$x_i=\frac{v_i}{2}+\sum_{k, \pi(k)<\pi(i)}v_k$}; that is each vertex is positioned at its center of mass, capturing a segment on the real axis that equals its length. The original form of the problem is the special case where all the volumes are equal. In particular, we would like to concentrate on the minimum linear arrangement (where $p=1$) and the minimum 2-sum problem (M2SP) that were proven to be NP-complete in \cite{gjs,georgepothen} and whose solution can serve as an approximation for many different linear ordering problems replacing the spectral approaches \cite{safro2004,safro2005}.
%In particular, we would like to concentrate on the minimum bandwidth problem which seeks a linear layout that minimizes the maximal stretched edge, i.e., $bw(G)=\min_{\pi}\max_{ij}w_{ij}|\pi(i)-\pi(j)|$. The minimization functional of the bandwidth problem for unweighted graph ($w_{ij}=1,~\forall ij\in E$) can be formulated in term of $\sigma_p(G, \pi)$ :
%\begin{equation}
%bw(G, \pi)=\lim_{p\rightarrow \infty}(\sigma_p(G, \pi))^{1/p}~~~,\label{banddef}
%\end{equation}
%since $\sigma_p(G,\pi)$ for large enough $p$ is practically dominated by the longest edge, i.e., by the bandwidth \cite{juvan}. Besides the minimum bandwidth problem there are two NP-hard well known problems defined by M$p$SP: (a) the minimum linear arrangement problem (where $p=1$, see \cite{survey:petit}) and (b) the minimum $2$-sum problem (where $p=2$, see \cite{georgepothen}).
\subsection{Partitioning}
\par The goal of the 2-partitioning problem is to find a partition of $V$ into two disjoint nonempty subsets $\Pi_1$ and $\Pi_2$ such that
\begin{equation}
\label{part-def}
    \text{minimize }  \sum_{i\in \Pi_1 , j\in \Pi_2}w_{ij}~~,~~\\
    \text{subject to} ~~ |\Pi_k|\leq (1+\alpha)\cdot \frac{|V|}{2}~~,
     ~(k=1,2) ~~,
\end{equation}
where $\alpha$ is a given {\it imbalance factor}.
\par Graph partitioning is an NP-hard problem \cite{Garey79} used in many fields of
computer science and engineering. Applications include VLSI
design, minimizing the cost of data distribution in parallel
computing and optimal tasks scheduling. Because of its practical
importance, many different heuristics (spectral~\cite{posili90}, combinatorial~\cite{keli70,fima82},
evolutionist~\cite{buiMoon96}, etc.) have been developed to
provide an approximation in a reasonable (and, one hopes, linear)
computational time. However, only the introduction of multilevel
methods for partitioning
\cite{metis,webscotch,alpert97multilevel,MeyerhenkeMonienSauerwald08new,Walshaw-AoOR-04,evo03proc,doritpart,barnard94fast,hendrickson95multilevel,kaku95a,Abou-RjeiliK06}
has really provided a breakthrough in efficiency and quality.
%\par The superscript in a matrix (vector) notation $X^{(k)}$  ($x^{(k)}$)will correspond to the iteration number. The bold superscript index in a vector notation $x^{\bf l}$ will correspond to an $l$-th column of matrix $X$.

\section{The coarsening algorithm}\label{algorithm}
\par In the multilevel framework a hierarchy of decreasing size graphs
$G_0,G_1,...,G_k$ is constructed. Starting from the given graph,
$G_0 = G$, we create by recursive {\it coarsening} the sequence $G_1,...,G_k$,
then solve the coarsest level directly, and uncoarsen the
solution back to $G$.
%This entire process is called a V-cycle.
\par In general, the AMG-based coarsening is interpreted as a process
of weighted aggregation of the graph nodes to define the nodes of
the next coarser graph. In weighted aggregation each node can be
divided into fractions, and different fractions belong to
different aggregates. The construction of a coarse graph from a
given one is divided into three stages. First a subset of the fine
nodes is chosen to serve as the {\it seeds} of the aggregates (the
nodes of the coarse graph). Then the rules for aggregation are
determined, thereby establishing the fraction of each nonseed
node belonging to each aggregate. Finally, the graph couplings
(or edges) between the coarse nodes are calculated. The entire
coarsening scheme is shown in Algorithm \ref{alg-gen-coarsening}.

\par The AMG-based multilevel framework for graph optimization problems is discussed, for example, in \cite{safro2005}.
In the present work we generalize the coarsening part of the AMG-based
framework. The problem-dependent solution of the coarsest level
and the uncoarsening are not changed here. They are
fully described in \cite{safro2005} and references therein.
\par The principal difference between the previous AMG-based coarsening approaches
\cite{safro2005,Hu01amultilevel,cheval-mlpartcompar}
and the new {\it relaxation-based} approach is the improved
measure, the {\it algebraic coupling},
assigned to each edge, or, more generally, between any two nodes, in the graph. The algebraic
coupling is the reciprocal of the calculated {\it algebraic distance}
introduced below.
\par
\begin{figure}[h]
\centering
\includegraphics[width=6cm]{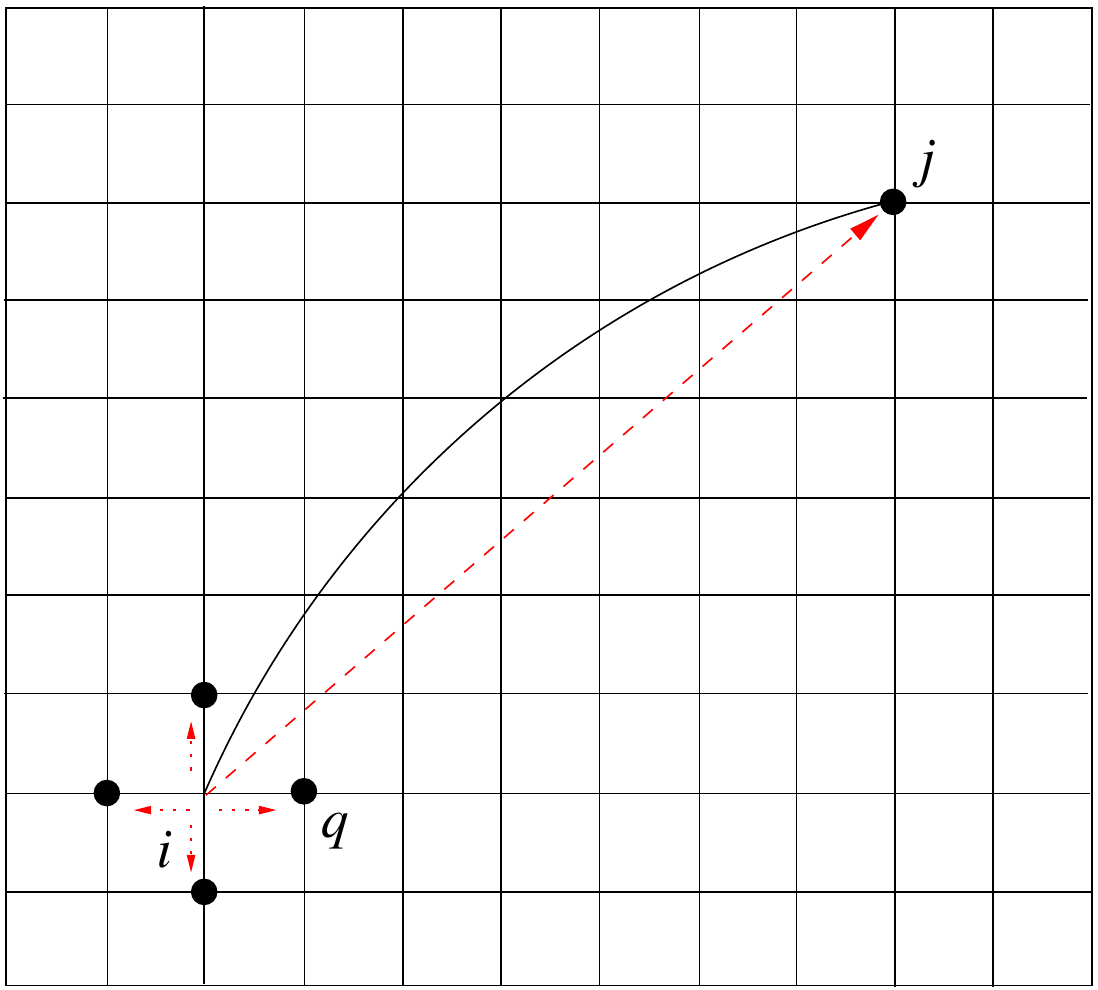}
\caption{Mesh graph with an additional edge between nodes $i$
and $j$. The black dots mark some of the nodes selected to serve
as the seeds of the coarse aggregates; see Algorithm \ref{alg-coarse-nodes}.}\label{fig:meshexample}
\end{figure}

 {\bf Algebraic distance and coupling.} The need for an improved
measure for the graph couplings can be explained by observing the
graph depicted in Figure \ref{fig:meshexample}: one additional edge
$ij$ (connecting nodes $i$ and $j$) is added to a regular twodimensional mesh. While coarsening, nodes $i$
and $j$ clearly should not belong to the same aggregate unless their coupling
is much stronger than other graph couplings. However, if the
weight of $ij$ is just somewhat larger than all other graph edges, and if the
black dots are some of the seeds of the coarse aggregates (chosen
by some AMG-based criterion; see, for example, Algorithm \ref{alg-coarse-nodes}), node $i$ will
tend to be aggregated with node $j$, rather than with any of its neighbors.
Such a decision will create bad coarse-level approximations
in many optimization problems (e.g., linear ordering and
partitioning). Moreover, at the next-coarser levels the approximation
may further deteriorate by making similar wrong decisions, making the entire neighborhood of $i$ close to $j$, thereby promoting linear arrangements in which {\it many} local couplings would unnecessarily become long-range ones. To
prevent this situation we would like to have a measure that not only
evaluates the coupling between $i$ and $j$ according to
the {\it direct} coupling between them but also takes into
account the contribution of connections between the {\it
neighborhoods} of $i$ and $j$. That is, if the immediate (graph)
neighbors of $i$ are connected to those of $j$,
the coupling between $i$ and $j$ should be enhanced, while if
$i$'s neighbors are not connected to those of $j$, as in Figure
\ref{fig:meshexample}, a significant weakening of the
$ij$ coupling is due. This
measure will prevent possible errors while coarsening.
%The main
%use here will be to detect those graph couplings which are much
%weakened when considering their algebraic coupling and forbid
%merging together the nodes at their endpoints.%\section{Algebraic distance}
%\par Let us start to introduce the notion of algebraic distance from a simple example which
%demonstrates when an AMG-based coarsening can fail. Imagine a mesh graph with one additional
%diagonal $ij$ as depicted in Figure \ref{fig:meshexample}.
%All the edges except the diagonal have weight 1 while the diagonal 2.
% Let the black small circles represent the centers of aggregates chosen by some AMG-based coarsening.
% Easy to see that during a coarse aggregate formation with interpolation order 1,
% seed $j$ will attract fine vertex $i$.
% Such decision can create a local conflict in many optimization problems
% (e.g. linear ordering and partitioning). Moreover, at the next coarse levels this conflict
% will be reinforced by making similar local conflict decisions that are based on the previous
% level fine-to-coarse attractions. The concept of algebraic distance is created to prevent these conflicts.

\par We introduce the notion of {\it algebraic distance}, which is
based on the same set of {\it test vectors} (TVs) being used in the
bootstrap AMG (BAMG) \cite{amg}. The key new ingredient of the
adaptive BAMG setup is the use of several TVs, collectively
representing algebraically smooth error, to define the
interpolation weights. When a priori knowledge of the nature of
this error is not available, slightly relaxed random vectors are
used for this task. A set of some $K$ low-residual TVs
$\{x^{(k)}\}^{K}_{k=1}$ can first be obtained by relaxation.
Namely, each $x^{(k)}$ is a result of $r$ fine-level
relaxation sweeps on the homogeneous equation $Ax=0$, starting
from a {\it random} approximation, where $A$ is the Laplacian of
the graph. In particular, we have used a small number (usually r=10)
of Jacobi under relaxation sweeps with $\omega =0.5$. That is, the
new value for each $x^{(k)}$, $k=1,...,K$ (in our tests $K=20$) is
\begin{equation}\label{sor-iter}
x^{(k)}_{NEW} = (1-\omega)x^{(k)} + \omega x_{JAC}^{(k)}~,
\end{equation}
where
\begin{equation}\label{jacobi-iter}
x_{JAC}^{(k)} = D^{-1}(D-A)x^{(k)} ~,~
\end{equation}
$D$ being the diagonal of $A$. The
algebraic distance from node $i$ to node $j$ is defined over the
$K$ relaxed TVs by
\begin{equation}\label{alg-dis}
d_{ij} = \max_{k=1}^K |x^{(k)}_i - x^{(k)}_j|~.
\end{equation}
Other definitions, such as
\begin{equation}\label{other-alg-dis}
d_{ij} = \sum_{k=1}^K (x^{(k)}_i - x^{(k)}_j)^2
\end{equation}
are also possible.
Hence, {\it only} if $d_{ij}$ is small may nodes $i$ and $j$ be
aggregated into the same coarse node. The algebraic coupling
between $i$ and $j$, $c_{ij}$, is defined as the reciprocal of
$d_{ij}$:
\begin{equation}\label{alg-coupling}
c_{ij} = 1/d_{ij}~.
\end{equation}
\begin{algorithm}
\SetLine \KwData{$Q$ , $\nu$} \KwResult{coarse graph}

For every edge $ij$ derive its algebraic distance $d_{ij}$
(\ref{alg-dis}) or (\ref{other-alg-dis}) and algebraic coupling $c_{ij}$
(\ref{alg-coupling})\; SelectCoarseNodes($Q$ , $\nu$)\; Define the
coarse graph using the matrix $P$ in Equation (\ref{interp-mat})\;
\caption{Coarsening scheme}\label{alg-gen-coarsening}
\end{algorithm}

\par We return to the example in Figure \ref{fig:meshexample} and demonstrate the
outcome of Definition (\ref{alg-dis}) by comparing $d_{ij}$ with $d_{i*} = $ min$\{d_{is}|s$ a nearest neighbor of $i\}$.
We show that $i$ will not tend to be connected to $j$ unless $w_{ij}$ equals the sum of $i$'s other couplings. Furthermore, we show that even if $i$ {\it is} connected to $j$ as a result of strong $w_{ij}$, $i$'s other neighbors will not tend to be connected to $i$ as well but will prefer other neighbors; hence the neighborhoods of $i$ and $j$ will not tend to be connected to each other.
Consider Table
\ref{tab:meshexampleDiag}. The number $K$ of TVs is given in the leftmost column.
The number $r$ of Jacobi relaxation sweeps
varies from $10$ to $100$ as shown in the second to the left column. Each of
the four columns to the right presents the (natural) logarithm of $d_{ij}/d_{i*}$,
averaged over 100 independent runs, for graph
couplings $w_{uv}=1$ when $u$ and $v$ are nearest neighbors, and
$w_{ij}=1,2,3,$ or 4 as shown. The numbers in parentheses are the corresponding standard deviations.
Clearly the strength of the coupling between $i$
and $j$ is relatively decreased when measured by the algebraic distance. For
instance, if the graph coupling between $i$ and $j$ is 1 (as are
all other couplings in the graph), then after 20 relaxation sweeps (with $K=10$)
$d_{ij}$ is three times bigger than the minimum of the
(algebraic distance of the) edges to $i$'s four nearest neighbors. Thus,
 the algebraic coupling between $i$ and $j$ is
{\it not} the strongest coupling of $i$ (not even close to it),
and hence it is guaranteed that $i$ and $j$ will not belong to the same coarse node.
The importance of using more than just 1 TV
can be seen from the values of the standard deviations: The use of 1 TV results in standard deviations similar to the average, which means that $ln(d_{ij}/d_{i*})$ has
a significant chance to become negative, so $ij$ has a significant chance to be the strongest
coupling of $i$. With 10 TVs this chance becomes much smaller, at least for $w_{ij} \leq 2$.
Even with 10 TVs, however, the chance grows with
%of almost the calculated values (especially for larger $w_{ij}$),
%which indicates how imprecise are the
$w_{ij}$, becoming more than $50\%$ roughly when $w_{ij} \geq 4$.
Thus, the aggregation of $i$ with $j$  becomes likely. This by itself is fine and justified.
What we really need to avoid is that entire neighborhoods of $i$ and $j$ will, as a result, be aggregated at some coarser level.
%measurements, with more TVs the standard deviations are relatively reduced which means that
%the distances measured for $ij$ and for other couplings of $i$ and $j$ become distinguishable. Finally,
%when the strength of $w_{ij}$ grows, its ratio to the other couplings of $i$ decreases, as its importance becomes evident.
%When $w_{ij}=4$ the ratio may even become negative which means that $i$ will occasionally be connected to $j$.
Hence, it is important to see whether the neighbors of $i$ will tend to be aggregated with $i$ (and thus also with $j$) or will prefer their other neighbors. To see that, we calculate the (natural) logarithm of $d_{qi}/d_{q^i*}$, where
$d_{q^i*} = $ min$\{d_{qs}|s$ a nearest neighbor of $q$ other than $i\}$.
%averaged over 100 independent runs, for graph
%couplings $w_{uv}=1$ when $u$ and $v$ are nearest neighbors, and
%$w_{ij}=1,2,3$ or 4 as shown.
%$ln$ of the ratio of some $k$, a neighbor of $i$ to the minimum algebraic distance of $k$'s other (than $i$) neighbors.
As shown in Table \ref{tab:meshexampleNonDiag}, $q$ would rather be aggregated
with one of its other-than-$i$ neighbors. For example, for $K=10$, $r=20$ and $w_{ij}=4$ out of the 100 runs, in 95 $q$ would have been connected with $i$.
The main conclusion is that nodes $i$ and $j$ do not tend to be connected as long as $w_{ij}$ is smaller than the sum of all other couplings of $i$ or of $j$. When the coupling is of the same strength, they will be connected about half the time, but then, not less important, the neighbors of $i$ (and similarly of $j$) will {\it not} tend to join them but will prefer to be connected to other nearest neighbors nodes.
Similar results are obtained when using (\ref{other-alg-dis}) to calculate $d_{ij}$.
\par With the notion of the algebraic coupling in mind, the
coarse nodes selection and the calculation of the aggregation
weights are modified as follows.
%The construction of the coarse graph remains unchanged.

\begin{table}[h]
\begin{center}
{
\begin{tabular}{|c|c|cccc|}
\hline
& & \multicolumn{4}{c|}{$w_{uv}=1$ for $(u,v)$ nearest neighbors} \\ \cline{3-6}
$K$ & $r$ & $w_{ij}=1$ & $w_{ij}=2$ & $w_{ij}=3$ & $w_{ij}=4$ \\
\hline
\multicolumn{1}{|c|}{\multirow{4}{*}{1}} &
\multicolumn{1}{|c|}{10} & 2.47(1.51) & 1.88(1.74) & 1.38(1.85) & 1.14(1.69)\\
\multicolumn{1}{|c|}{}                        &
\multicolumn{1}{|c|}{20} & 2.74(1.74) & 2.1(1.59) & 1.4(1.26) & 1.44(1.59) \\
\multicolumn{1}{|c|}{}                        &
\multicolumn{1}{|c|}{50} & 2.65(1.36) & 1.98(1.76) & 1.92(1.41) & 1.5(1.59) \\
\multicolumn{1}{|c|}{}                        &
\multicolumn{1}{|c|}{100}& 3.03(1.72) & 2.14(1.32) & 1.51(1.42) & 1.16(1.78) \\
\hline
\multicolumn{1}{|c|}{\multirow{4}{*}{5}} &
\multicolumn{1}{|c|}{10} & 1(0.502) & 0.628(0.416) & 0.24(0.417) & -0.0484(0.397)\\
\multicolumn{1}{|c|}{}                        &
\multicolumn{1}{|c|}{20} & 1.34(0.442) & 0.825(0.415) & 0.435(0.358) & 0.208(0.332) \\
\multicolumn{1}{|c|}{}                        &
\multicolumn{1}{|c|}{50} & 1.68(0.342) & 1.04(0.338) & 0.643(0.306) & 0.362(0.296) \\
\multicolumn{1}{|c|}{}                        &
\multicolumn{1}{|c|}{100} &1.78(0.467) & 1.06(0.392) & 0.743(0.369) & 0.396(0.359) \\
\hline
\multicolumn{1}{|c|}{\multirow{4}{*}{10}} &
\multicolumn{1}{|c|}{10} & 0.821(0.281) & 0.443(0.294) & 0.022(0.293) & -0.244(0.313) \\
\multicolumn{1}{|c|}{}                        &
\multicolumn{1}{|c|}{20} &1.09(0.268) & 0.624(0.239) & 0.298(0.235) & 0.0126(0.253) \\
\multicolumn{1}{|c|}{}                        &
\multicolumn{1}{|c|}{50} & 1.49(0.263) & 0.86(0.235) & 0.504(0.226) & 0.2(0.204) \\
\multicolumn{1}{|c|}{}                        &
\multicolumn{1}{|c|}{100} & 1.69(0.315) & 1.01(0.275) & 0.572(0.264) & 0.285(0.271) \\
\hline
\end{tabular}
}

\caption{Statistical results (over 100 runs) for the average (and, in parentheses,
the standard deviation) of $ln(d_{ij}/d_{i*})$, calculated with
$K$ TVs and $r$ Jacobi relaxation sweeps for different relative strengths of $w_{ij}$.}
%$ln$ of the ratio between the algebraic distance of the edge $ij$ (in Figure
%\ref{fig:meshexample}) and the minimum among the algebraic distances of the
%four nearest neighbors (on the mesh) of $i$,
%performed for calculating the algebraic
%distance (\ref{alg-dis}). The top row shows
%graph coupling of the edge $ij$. All other edges are
%of equal weight 1. The number of random vectors $K=1$. All the
%values are rations between $d_{ij}$ and the average over $d_{uv}$
%for $uv \neq ij$.}
%$\frac{\rho_{ij}}{\text{avg}_{uv\neqij}\rho_{uv}}$.}
\label{tab:meshexampleDiag}
\end{center}
\end{table}

\begin{table}[h]
\begin{center}
%\small
{
\begin{tabular}{|c|c|cccc|}
\hline
& & \multicolumn{4}{c|}{$w_{uv}=1$ for $(u,v)$ nearest neighbors} \\ \cline{3-6}
$K$ & $r$ & $w_{ij}=1$ & $w_{ij}=2$ & $w_{ij}=3$ & $w_{ij}=4$\\
\hline
\multicolumn{1}{|c|}{\multirow{4}{*}{1}} &
\multicolumn{1}{|c|}{10} & 0.975(1.67) & 0.939(1.7) & 1.14(1.63) & 1.07(1.89)\\
\multicolumn{1}{|c|}{}                        &
\multicolumn{1}{|c|}{20} & 0.911(1.62) & 1.09(1.31) & 1.02(1.46) & 0.931(1.64) \\
\multicolumn{1}{|c|}{}                        &
\multicolumn{1}{|c|}{50} & 1.37(1.77) & 1.14(1.79) & 1.28(1.49) & 1.24(1.45) \\
\multicolumn{1}{|c|}{}                        &
\multicolumn{1}{|c|}{100} & 0.897(1.55) & 1.23(1.45) & 1.29(1.53) & 1.31(1.44) \\
\hline
\multicolumn{1}{|c|}{\multirow{4}{*}{5}} &
\multicolumn{1}{|c|}{10} & 0.382(0.534) & 0.482(0.428) & 0.416(0.52) & 0.587(0.487) \\
\multicolumn{1}{|c|}{}                        &
\multicolumn{1}{|c|}{20} & 0.434(0.444) & 0.472(0.366) & 0.592(0.486) & 0.663(0.458) \\
\multicolumn{1}{|c|}{}                        &
\multicolumn{1}{|c|}{50} &   0.498(0.436) & 0.755(0.53) & 0.784(0.526) & 0.813(0.455) \\
\multicolumn{1}{|c|}{}                        &
\multicolumn{1}{|c|}{100} & 0.501(0.522) & 0.746(0.544) & 0.812(0.549) & 0.816(0.535) \\
\hline
\multicolumn{1}{|c|}{\multirow{4}{*}{10}} &
\multicolumn{1}{|c|}{10} & 0.283(0.312) & 0.299(0.316) & 0.376(0.307) & 0.401(0.357) \\
\multicolumn{1}{|c|}{}                        &
\multicolumn{1}{|c|}{20} & 0.362(0.281) & 0.419(0.295) & 0.449(0.288) & 0.441(0.327) \\
\multicolumn{1}{|c|}{}                        &
\multicolumn{1}{|c|}{50} & 0.448(0.311) & 0.531(0.35) & 0.672(0.351) & 0.604(0.333) \\
\multicolumn{1}{|c|}{}                        &
\multicolumn{1}{|c|}{100} & 0.464(0.377) & 0.682(0.348) & 0.839(0.374) & 0.749(0.39) \\
\hline
\end{tabular}
}
\caption{Statistical results (over 100 runs) for the average (and, in parentheses,
the standard deviation) of $ln(d_{qi}/d_{q^i*})$ (see Figure
\ref{fig:meshexample}), calculated with
$K$ TVs and $r$ Jacobi relaxation sweeps for different relative strengths of $w_{ij}$.}
%Statistical results (over 100 runs) for the logarithm of the ratio between the algebraic distance of the edge
%connecting $i$ to one of its nearest neighbors, say $q$ (see Figure
%\ref{fig:meshexample}) and with the minimum of the algebraic distances among the
%three other nearest neighbors (on the mesh) of $q$, $d_{qi}/d_{q^i*}$,
%employing $K$ TVs and $r$ Jacobi relaxation sweeps for different strengths of $w_{ij}$.}
%performed for calculating the algebraic
%distance (\ref{alg-dis}). The top row shows
%graph coupling of the edge $ij$. All other edges are
%of equal weight 1. The number of random vectors $K=1$. All the
%values are rations between $d_{ij}$ and the average over $d_{uv}$
%for $uv \neq ij$.}
%$\frac{\rho_{ij}}{\text{avg}_{uv\neqij}\rho_{uv}}$.}
\label{tab:meshexampleNonDiag}
\end{center}
\end{table}

%$\rho$-based lies in the usage of edge weight as a basic  argument
%of the relationships between edge's endpoints. In almost all
%places where the edge weight was a part of some calculation or
%decision, in the $\rho$-based coarsening, the algebraic distance
%$\rho$ replaces the edge weight by considering the influence of
%the neighborhoods of the endpoints.
%\par In general, like a classical AMG coarsening, the $\rho$-based coarsening is interpreted as a process of weighted aggregation of the graph nodes to define the nodes of the next coarser graph. In weighted aggregation each node can be divided into fractions, and different fractions belong to different aggregates. The construction of a coarse graph from a given one is divided into three stages: first a subset of the fine nodes is chosen to serve as the seeds of the aggregates (the nodes of the coarse graph), then the rules for
%interpolation are determined, thereby establishing the fraction of each non-seed node belonging to each aggregate, and finally the strength of the connections (or edges) between the coarse nodes is calculated. The entire coarsening scheme is shown in Algorithm \ref{alg-gen-coarsening}.
%\par {\bf Coarse nodes.} The selection of the set of coarse nodes
%is basically the same as in \cite{safro2005}, with the exception that
%it relies not on the graph couplings $w_{ij}$'s but on the
%algebraic couplings $c_{ij}$'s defined in (\ref{alg-coupling}).
%However, for completeness, we repeat its description.
\par {\bf Seed selection.} The construction of the set of seeds $C$ and its complement
$F$ is guided by the principle that each $F$-node should be
``strongly coupled" to $C$. We will include in $C$ nodes with
exceptionally large volume or nodes expected (if used as seeds)
to aggregate around them an exceptionally large total volume of $F$-nodes.
We start with $C = \emptyset$, hence $F=V$, and then sequentially
transfer nodes from $F$ to $C$, as follows. As a measure of how
large an aggregate seeded by $i \in F$ might grow, we define its
%\par Like in the classical AMG, the construction of
%coarse nodes begins with the division of $V$ into two disjoint
%parts: $C$ - set of seeds, and its complement $F$. This division
%is guided by the principle that each $F$-node should be "strongly
%coupled" to $C$. However, in the $\rho$-based coarsening this
%principle will be based strictly on the algebraic distances.
%Starting with $C=\emptyset$ and $F=V$, we sequentially transfer
%nodes from $F$ to $C$, considering as a measure of a future
%aggregate size seeded by $i\in F$, the redefined
{\it future volume} $\vartheta_i$
%from \cite{safro2005}
by
\begin{equation}\label{future-vol}
\vartheta_i = v_i + \sum_{ij\in E}v_j\cdot
\frac{c_{ji}}{\sum_{jk\in E}c_{jk}}~~.
\end{equation}
%where $N(i)$ is the list of all nodes directly connected to $i$ in
%the graph.
Nodes with future volume larger than $\nu$ times the
average of the $\vartheta_i$'s are first transferred to $C$ as
most ``representative" (in our tests $\nu=2$). The insertion of
additional $F$-nodes to $C$ depends on a ``strength of coupling to $C$" threshold $Q$ (in our
tests $Q=0.5$),
% The process terminates when all nodes are either
%in $C$ or "strongly coupled" to $C$
as specified in Algorithm
\ref{alg-coarse-nodes}.
\begin{algorithm}
\SetLine
\KwData{$Q$, $\nu$}
\KwResult{set of seeds $C$}

$C\leftarrow \emptyset,~F\leftarrow V$\; Calculate $\vartheta_i$
(\ref{future-vol}) for each $i\in F$, and their average
$\overline{\vartheta}$\; $C\leftarrow$nodes $i$ with $\vartheta_i
> \nu \cdot \overline{\vartheta}$\; $F\leftarrow V\setminus C$\;
\SetLine \ForAll {$i\in F$ in descending order of $\vartheta_i$} {
\SetNoline \lIf{$( \sum_{j\in (C \cap N(i))} c_{ij} / \sum_{j\in N(i)} c_{ij} )
\leq Q$ or $( \sum_{j\in (C \cap N(i))} w_{ij} / \sum_{j\in N(i)} w_{ij} )
\leq Q$}{ move $i$ from $F$ to $C$}\; }
\caption{SelectCoarseNodes($Q$, $\nu$)}\label{alg-coarse-nodes}
\end{algorithm}
\par {\bf Coarse nodes.} Each node in the chosen set $C$ becomes
the seed of an aggregate that will constitute one coarse-level
node. Next it is necessary to determine for each $i \in F$ a list
of $j \in C$ to which $i$ will belong. Define {\it caliber}, $l$,
to be the maximal number of $C$-points allowed in that list. The
selection we propose here is based on both measures: the graph
couplings $w_{ij}$'s and the algebraic couplings $c_{ij}$'s.
Define for each $i \in F$ a coarse neighborhood
$\bar{\bar{N}}^C(i)=\{ j \in C : ij \in E \}$. Set $D$ to be the
maximal $c_{ij}$ in $\bar{\bar{N}}^C(i)$. Construct a possibly
smaller coarse neighborhood by including only nodes with strong
algebraic coupling ${\bar{N}}^C(i)=\{ j \in \bar{\bar{N}}^C(i) :
c_{ij} \geq \beta*D \}$, we use $\beta=0.5$. If ${\bar{N}}^C(i) > l$, then the final coarse
neighborhood ${N}^C(i)$ will include the first $l$ largest
$w_{ij}$'s in ${\bar{N}}^C(i)$. If ${\bar{N}}^C(i) \leq l$, then
$N^C(i) \leftarrow {\bar{N}}^C(i)$. This construction of the coarse
neighborhood $N^C(i)$ of $i \in F$ is summarized in Algorithm
\ref{alg-interpsort2}. (In the results below we have used only $l=1$ and $l=2$.)
The classical AMG interpolation matrix
%is the
%most important ingredient of the multilevel framework that has to
%be improved by replacing the local weights $w_{ij}$ by
%$\rho'_{ij}$ which better reflects the connectivity of two
%endpoints. The $\rho$-based interpolation matrix
$P$ (of size $|V|\times |C|$) is then defined by
\begin{equation}\label{interp-mat}
P_{ij}~=~ \left\{
  \begin{array}{lll}
  w_{ij}/\sum\limits_{k\in N^C(i)}w_{ik} & \textsf{for }i\in F , ~ j\in N^C(i) \\
  1               & \textsf{for }i\in C, ~ j=i \\
  0               & \textsf{otherwise} ~~~~~~.
  \end{array} \right.
\end{equation}
$P_{ij}$ represents the fraction of $i$ that will belong to the
$j$th aggregate.

\begin{algorithm}
\SetLine \KwData{$l$, $i$, $\beta$} $\bar{\bar{N}}^{C}(i) \leftarrow \{ j \in
C~:~ ij \in E \}$\; $D~= \textsf{max}_{j \in \bar{\bar{N}}^{C}(i)}
c_{ij}$\; ${\bar{N}}^{C}(i)= \{ j \in \bar{\bar{N}}^{C}(i)  :
c_{ij} \geq \beta*D \}$\;
%Dilute $N^{\rho}(i)$ by throwing out $j\in
%N^{\rho}(i)$ with $\rho'_{ij} < \frac{\max_k \rho'_{ik}}{2}$\;
\SetVline
% \If{$\bar{\bar{N}}^{C}(i)$ was not diluted}{ Dilute
%$\bar{\bar{N}}^{C}(i)$ by throwing out $\lfloor
%|\bar{\bar{N}}^{C}(i)| \cdot
%\min(\frac{|\bar{\bar{N}}^{C}(i)|-c}{|\bar{\bar{N}}^{C}(i)|},\frac{1}{2})\rfloor$
%of nodes $j$ with small $d_{ij}$\; }
\If{$l<|\bar{N}^{C}(i)|$}{
$N^{C}(i) \leftarrow $ the $l$ largest $w_{ij}$'s in
$\bar{N}^{C}(i)$\; } \If{$l \geq |\bar{N}^{C}(i)|$}{ $N^{C}(i)
\leftarrow \bar{N}^{C}(i)$\; } \caption{The coarse neighborhood
$N^{C}(i)$}\label{alg-interpsort2}
\end{algorithm}

%where $N^{\rho}_i=\{j\in C~:~ j\text{ is not
%diluted by } \alpha(\beta,i) \}$ and $\alpha(\beta,i)$ is a
%control function for the coarse neighborhood size of $i$ bounded
%by interpolation order $\beta$. In the $\rho$-based scheme we
%suggest to design $\alpha$ using both algebraic distance and graph
%edge weight function as shown in Algorithm \ref{alg-interpsort}.
%\begin{algorithm}
%\SetLine
%\KwData{$\beta$, $i$}
%$N^{\rho}(i) \leftarrow \{j\in C~:~ ij\in E \}$\;
%\SetVline
%\If{$\beta<|N^{\rho}(i)|$}{
%Dilute $N^{\rho}(i)$ by throwing out $|N^{\rho}(i)| \cdot \min(\frac{|N^{\rho}(i)|-\beta}{|N^{\rho}(i)|},\frac{1}{2})$ of nodes $j$ with small $\rho'_{ij}$\;
%}
%\If{$\beta<|N^{\rho}(i)|$}{
%Dilute $N^{\rho}(i)$ by throwing out $|N^{\rho}(i)|-\beta$ nodes $j$ with small $w_{ij}$\;
%}
%\caption{$\alpha($interpolation order $\beta$, $F$-node $i$)}\label{alg-interpsort}
%\end{algorithm}
\par {\bf Coarse graph couplings.} The coarse couplings are constructed as follows. Let $I(k)$ be
the ordinal number in the coarse graph of the node that represents
the aggregate around a seed whose ordinal number at the fine level
is $k$. Following the weighted aggregation scheme used in
\cite{sharon}, the edge connecting two coarse aggregates, $p
=I(i)$ and $q=I(j)$, is assigned with the weight
$w_{pq}^{(coarse)}=\sum_{k\not= l}P_{ki}w_{kl}P_{lj}$. The volume
of the $i$th coarse aggregate is $\sum_j v_jP_{ji}$.
%\subsection{Some explanations in this subsection?}

\section{Computational results}\label{results}
\par We demonstrate the power of our new relaxation-based coarsening scheme by comparing its experimental
results with those of the
classical AMG-based coarsening
%In order to demonstrate the power
%of $\rho$-based coarsening we compared the experimental results
for three important NP-hard optimization problems: the minimum 2-sum
((\ref{minpsumdef}) with $p=2$), the minimum linear arrangement ((\ref{minpsumdef}) with $p=1$), and the minimum 2-partitioning (\ref{part-def}) problems. In all cases
%we compared
%the $\rho$-based coarsening with classical AMG-based coarsening by
the results are obtained by taking the lightest possible
uncoarsening schemes, so that differences due to the different coarsening schemes are least blurred.
\par We have implemented and tested the new coarsening scheme by
using the linear ordering packages developed in \cite{safro2004} and in \cite{safro2003} and the \scotch{} package \cite{webscotch} on a Linux
machine. The implementation is nonparallel and has not been
optimized. The results should be considered only qualitatively and
can certainly be improved by more advanced uncoarsening.
%Note that the primary goal of this work is to show the power of
%the new relaxation-based coarsening scheme compared with the old
%AMG-based version,
Thus, no intensive attempt to achieve the best-known
results for the particular test sets was done. The details
regarding the uncoarsening schemes for the above problems are
given in \cite{safro2004,safro2003,cheval-mlpartcompar}.
\subsection{The minimum $p$-sum problem}
\par We present the numerical comparison for two minimum $p$-sum problems: the minimum 2-sum problem and the minimum linear arrangement. For these problems we have designed a full relaxation-based
coarsening solver and evaluated it on a test set of 150 graphs of
different nature, size ($|V| \leq 5\cdot 10^6$ and $|E| \leq 10^7$)
and properties. The test graphs are taken from
\cite{davis} and from real-life network data such as social networks, power grids, and peer-to-peer connections. Our solvers are free and can be downloaded with
detailed solutions for every graph from \cite{safroproj}.
To emphasize the difference in the minimization results
between the two coarsening schemes (the relaxation-based and the classical AMG-based schemes), we measure the results
obtained at the end of the multilevel cycle {\it before} the final local
optimization postprocessing (Gauss-Seidel relaxation and the local processing in \cite{safro2004,safro2003}),
%Since this local optimization is
%very powerful (especially for the minimum 2-sum problem) we present comparison also
as well as after its application. Moreover, we use small calibers, $l=1,2$, since these demonstrate
more sharply the quality of matching between the $F$-points and
the $C$-points. For higher calibers it is also important to use
the adaptive BAMG scheme \cite{amg} for
calculating the interpolation weights, which is beyond the scope
of this work. Small calibers are important for maintaining the
low complexity of the multilevel framework, which is vital, for example, for hypergraphs and expanders.

\par
\begin{figure}[h]
\begin{minipage}[b]{0.5\linewidth} % A minipage that covers half the page
\centering
\includegraphics[width=7cm]{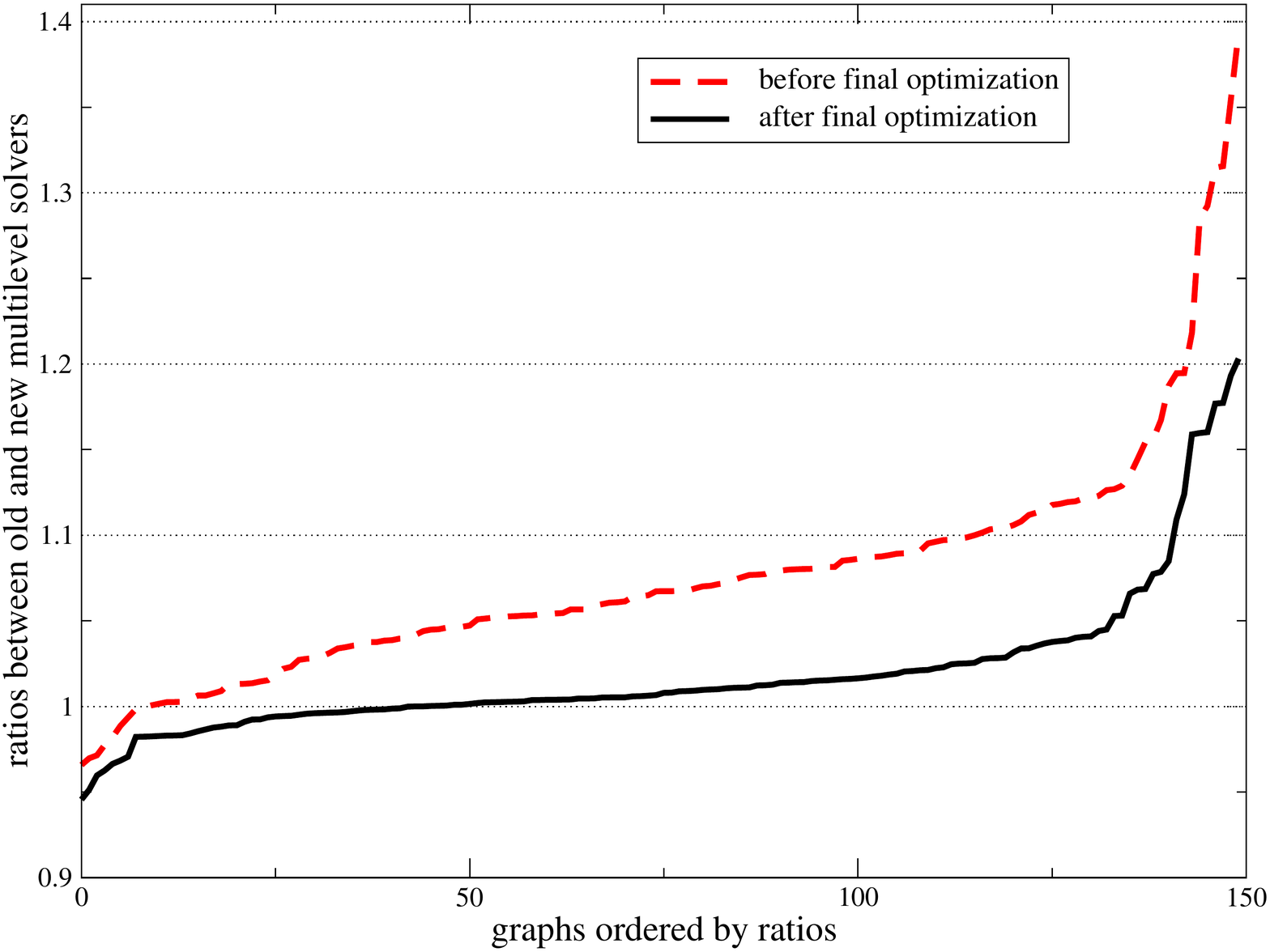}\\
(a) $l=1$
\end{minipage}
\hspace{0.5cm} % To get a little bit of space between the figures
\begin{minipage}[b]{0.5\linewidth}
\centering
\includegraphics[width=7cm]{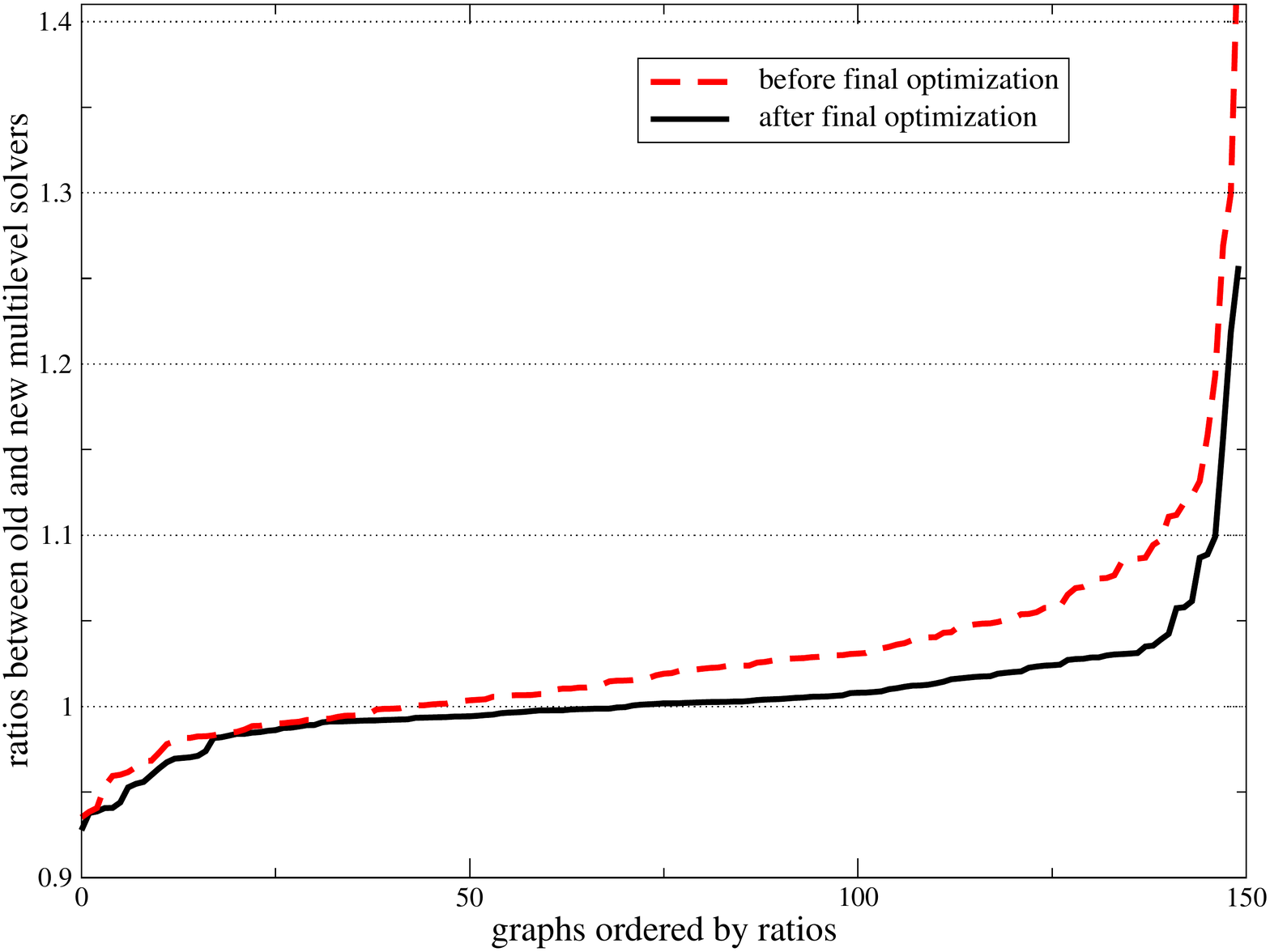}\\
(b) $l=2$
\end{minipage}
\caption{Results for the minimum 2-sum problem.}\label{fig:2sum}
\end{figure}

 {\bf The minimum 2-sum problem (M2SP).} A comparison of the relaxation-based and AMG-based
coarsenings with calibers 1 and 2 is presented in Figures
\ref{fig:2sum}(a) and \ref{fig:2sum}(b), respectively. Each x-axis
scale division corresponds to one graph from the test set. The
y-axis corresponds to the ratio between the average cost obtained by 100
runs of the AMG-based coarsening and the one obtained by 100 runs of
the relaxation-based coarsening. Each figure contains two curves:
the dashed curves with cost measurements before applying the postprocessing of local optimization (e.g., Gauss-Seidel
relaxation, Window Minimization \cite{safro2004}) and the regular curves with cost measurements
after adding such optimization steps.
%Gauss-Seidel and few iterations of Window Minimization
%postprocessing \cite{safro2004}.
Clearly most graphs benefit from
the relaxation-based coarsening, showing a ratio greater than 1. The ratio
decreases when more optimization is used, especially since the
Gauss-Seidel relaxation is powerful algorithmic component for this problem and thus
brings the results of the two coarsening schemes closer to each
other as is indicated by the regular curves. All the these results
were obtained with $K=10$ TVs. When we lowered $K$ to $5$, we observed no significant change in the results. Our number of
Jacobi overrelaxation sweeps $r=20$ cannot be reduced by more than twice
since this relaxation scheme is expected to evolve slowly. The detailed analysis of the convergence properties are presented in \cite{chen-safro-algdist-full}.
%\par WHAT DO WE OBSERVE? EXPLANATIONS? CONCLUSIONS?
% \begin{figure}
% \centering\includegraphics[width=11cm]{min2sum_io1_points}
% \caption{Results for the minimum 2-sum problem, $l=1$}\label{fig:2sum-1}
% \end{figure}
% \begin{figure}
% \centering\includegraphics[width=11cm]{min2sum_io2_points}
% \caption{Results for the minimum 2-sum problem, $l=2$}\label{fig:2sum-2}
% \end{figure}

\par {\bf The minimum linear arrangement problem.} Similarly to the previous problem, we designed a relaxation-based solver and established a series of experiments for the minimum linear arrangement problem. The experimental setup was identical to that of the M2SP. It was based on the solver designed in \cite{safro2003}. In this case we can observe even more significant improvement when employing the relaxation-based coarsening than for the M2SP.
\begin{figure}[h]
\begin{minipage}[b]{0.5\linewidth} % A minipage that covers half the page
\centering
\includegraphics[width=7cm]{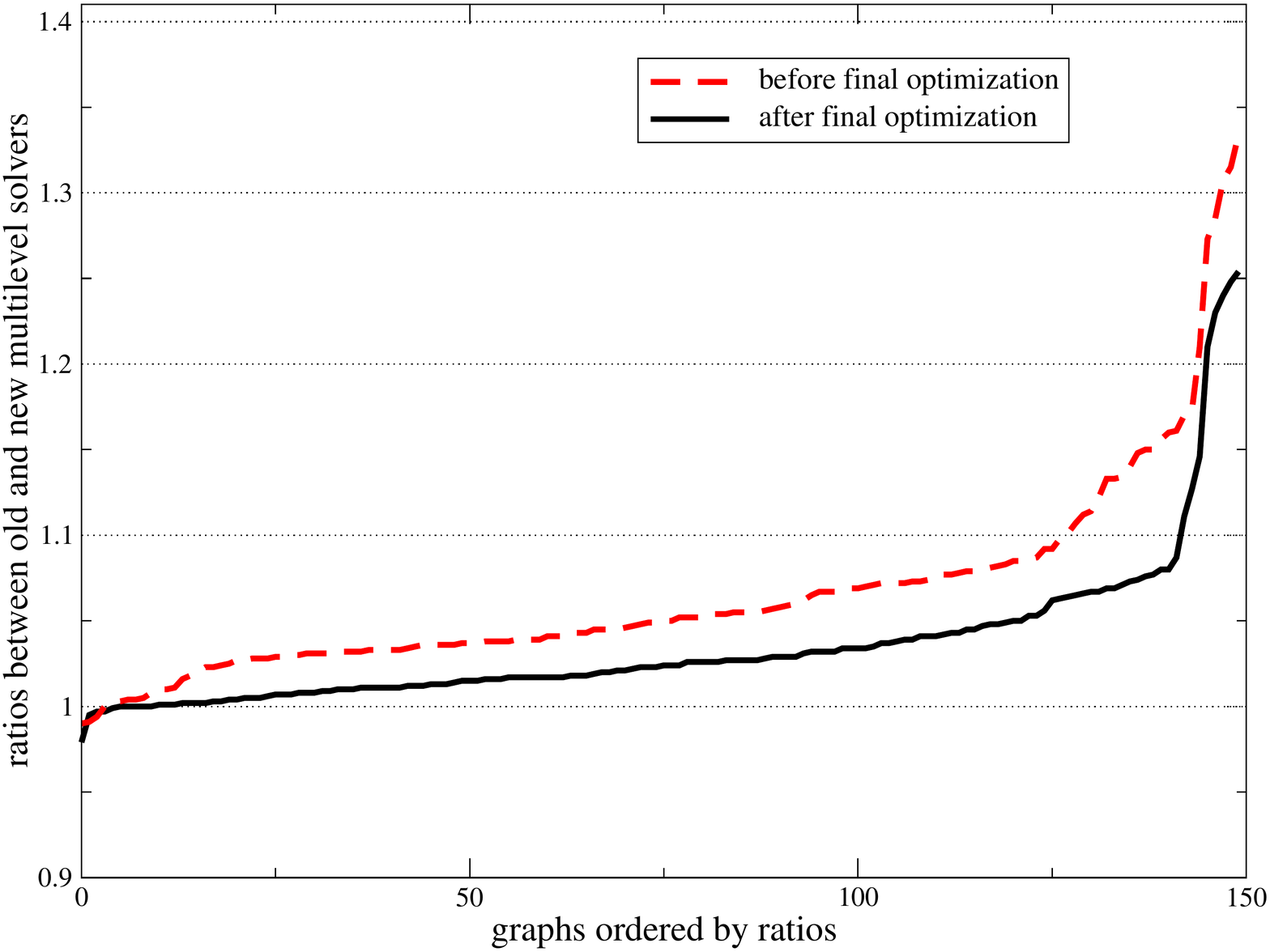}\\
(a) $l=1$
\end{minipage}
\hspace{0.5cm} % To get a little bit of space between the figures
\begin{minipage}[b]{0.5\linewidth}
\centering
\includegraphics[width=7cm]{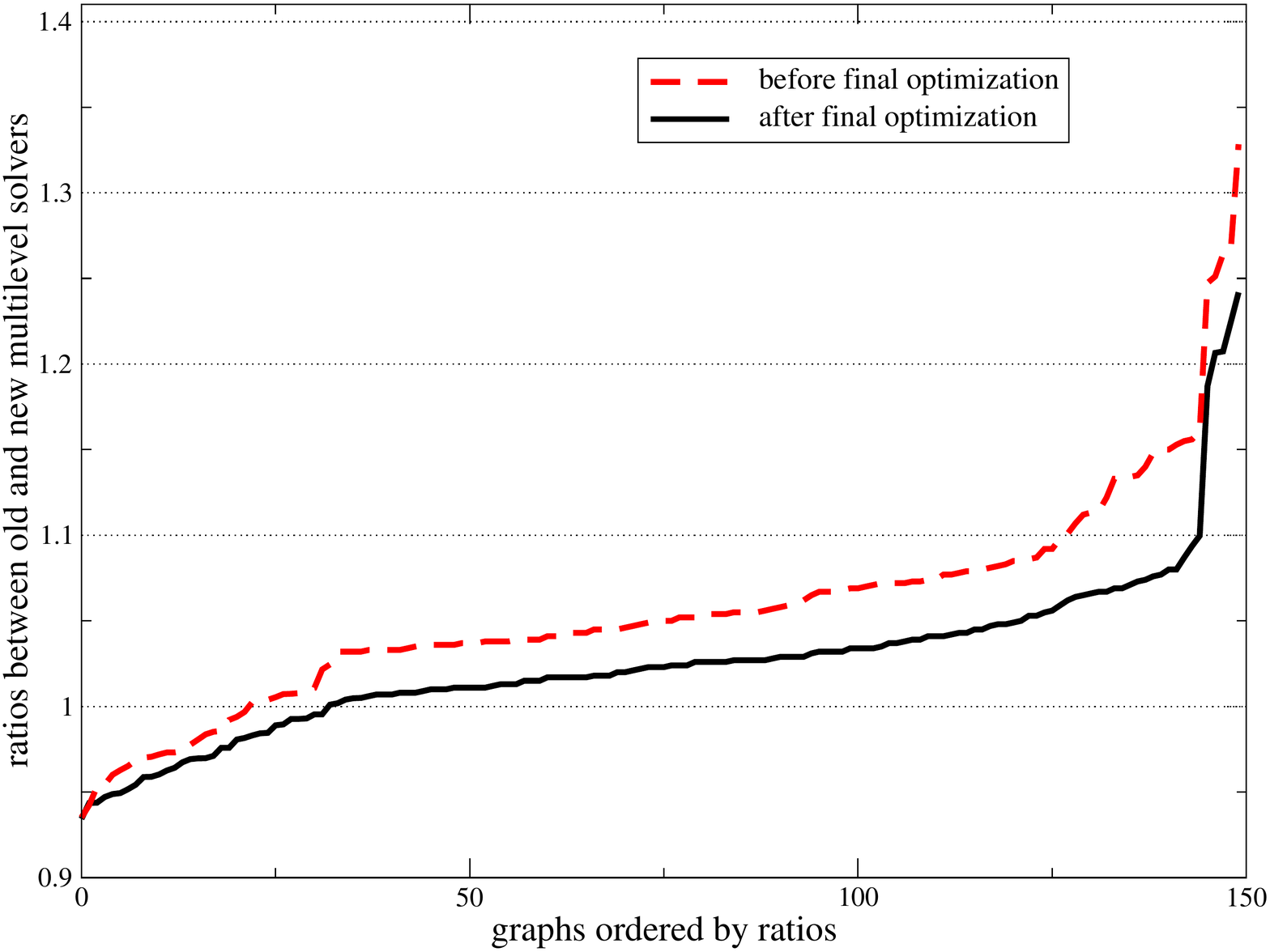}\\
(b) $l=2$
\end{minipage}
\caption{Results for the minimum 1-sum (linear arrangement) problem.}\label{fig:1sum}
\end{figure}
\par {\bf Which graphs are most beneficial?} It is remarkable that the most beneficial graphs in our test set come from VLSI design and general optimization problems. We know that these graphs are very irregular (compared, for example,   with finite-element graphs and with those that pose 2D/3D geometry). Thus, we may conclude that the algebraic couplings help to identify the weakness of nonlocal connections and prevent them from being aggregated. In several examples, we achieved the best known results with caliber 1, while using classical AMG-based approaches they can be achieved with bigger calibers only.
% \begin{figure}
% \centering\includegraphics[width=11cm]{min1sum-points-io1}
% \caption{Results for the minimum linear arrangement problem, $l=1$}\label{fig:1sum-1}
% \end{figure}
% \begin{figure}
% \centering\includegraphics[width=11cm]{min1sum-points-io2}
% \caption{Results for the minimum linear arrangement problem, $l=2$}\label{fig:1sum-2}
% \end{figure}

%\hspace{0.5cm} % To get a little bit of space between the figures
%\begin{minipage}[b]{0.5\linewidth}
%\centering
%\includegraphics[width=7cm]{min2sum_io2_points}
%(b) $l=2$
%\end{minipage}
%\caption{Results for the minimum 2-sum problem.}\label{fig:2sum}
%\end{figure}
\par
\begin{figure}[h]
\centering
\includegraphics[width=7cm]{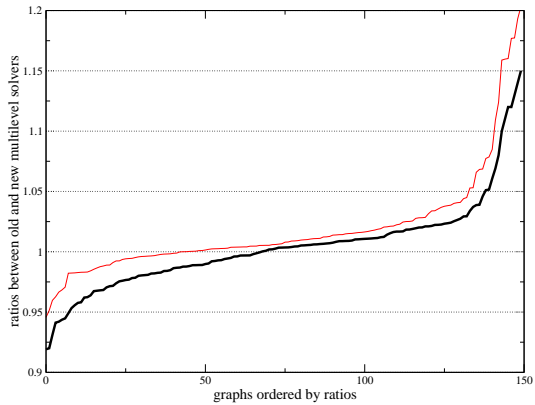}\\
\caption{Results for the minimum 2-sum problem. Comparison of the algebraic distance based only and mixed full relaxation based algorithms.}\label{fig:2sum-algdist-only}
\end{figure}

 {\bf An algebraic coupling-based algorithm.} We have also tried a straightforward algorithm in which, during coarsening, the weights of the graph are simply replaced by their algebraic couplings. That is, in the {\bf if} statement at the end of Algorithm \ref{alg-coarse-nodes}, only the first term is taken into account (namely,  $( \sum_{j\in (C \cap N(i))} c_{ij} / \sum_{j\in N(i)} c_{ij} )\leq Q$). Similarly, in Algorithm \ref{alg-interpsort2},
$w_{ij}$ (in the first {\bf if}) is replaced by $c_{ij}$.
%Another way to introduce the algebraic distance based considerations in the multilevel framework is to substitute completely with them all algorithmic components that decide according to the graph weights. In particular, we extracted the condition of the last ``if'' statement in Algorithm \ref{alg-coarse-nodes}, namely  $( \sum_{j\in (C \cap N(i))} w_{ij} / \sum_{j\in N(i)} w_{ij} )\leq Q$, and also the first ``if'' statement from Algorithm \ref{alg-interpsort2}.
We present the comparison of the obtained simple algebraic couplings based coarsening scheme with the mixed scheme described in Algorithms \ref{alg-coarse-nodes} and \ref{alg-interpsort2} in Figure \ref{fig:2sum-algdist-only}. The comparison was done for the M2SP including  postprocessing (of local optimization) using the same experimental setup. The bold curve corresponds to the ratios between the classical AMG-based results and the simple algebraic coupling-based coarsening scheme. To see the difference between this algorithm and the more elaborate one, we add a copy of its results, that is, the bold curve from Figure \ref{fig:2sum}(a). The mixed version is clearly better: in about 25 more graphs the results are improved. The average improvement was $1.5\%$.

\subsection{The minimum 2-partitioning problem}
%\par Graph partitioning is an NP-hard problem \cite{Garey79} used in many fields of
%computer science and engineering. Applications include VLSI
%design, minimizing the cost of data distribution in parallel
%computing, optimal tasks scheduling, etc. Because of its practical
%importance, many heuristics of different nature
%(spectral~\cite{posili90}, combinatorial~\cite{keli70,fima82},
%evolutionist~\cite{buiMoon96}, etc.) have been developed to
%provide an approximation in a reasonable (and, one hopes, linear)
%computational time. However, only the introduction of multilevel
%methods for partitioning
%\cite{metis,webscotch,alpert97multilevel,MeyerhenkeMonienSauerwald08new,Walshaw-AoOR-04,evo03proc,doritpart,barnard94fast,hendrickson95multilevel,kaku95a,Abou-RjeiliK06}
%has really provided a breakthrough in efficiency and quality.
\par We compared the relaxation-based coarsening and the classical AMG-based
 by combining two packages. The coarsening part was
 the same as in the minimum $p$-sum problems. The
uncoarsening was based on the \scotch{} package; details of
its fastest version can be taken from \cite{cheval-mlpartcompar}.
\par
\begin{figure}[h]
\centering
\includegraphics[width=7cm]{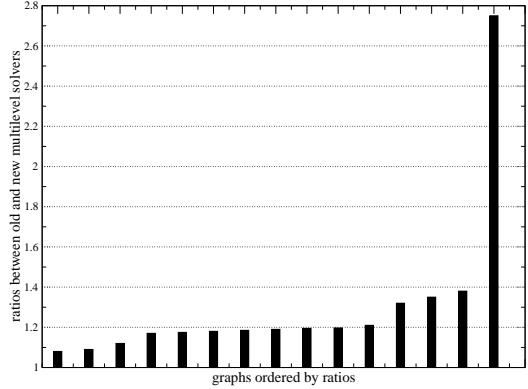}
\caption{Results for the minimum 2-partitioning problem.}\label{fig:partitioning}
\end{figure}

 The comparison of the relaxation-based and the AMG-based coarsenings with caliber
1 is presented in Figure \ref{fig:partitioning}. The
interpretation of x- and y-axes is similar to Figure
\ref{fig:2sum}.
%Since the implementation of the full
%relaxation-based solver is out of the scope of this work, the test
%set will be not very big.
Included are 15 graphs of different nature and size. The details
regarding the numerical results can be obtained from
\cite{safroproj}. The four best ratios are obtained for graphs with
power-law degree distributions.
% \begin{figure}[h]
% \centering
% \includegraphics[width=9cm]{partitioning.jpg}
% \caption{Results for the minimum 2-partitioning problem}\label{fig:partitioning}
% \end{figure}
%\subsection{Common observations}
%\begin{itemize}
%\item running time
%\item better performance on power low distributions
%\item confirmation of algdist goals on artificial graphs
%\end{itemize}
More results for the graph and hypergraph partitioning problems are reported in \cite{chen-safro-algdist-full}. Even though the algorithm used there only substitutes the original given couplings by their algebraic couplings, it is already clear that better results are observed for most tested instances of both graphs and hypergraphs.
%Both algorithms (with and without graph based considerations) are compared to the classical AMG-based coarsening.
%In contrast to the results obtained in several simple basic algorithms in \cite{chen-safro-algdist-full} where the graph considerations were just substituted with the algebraic distance, more sophisticated multilevel framework demonstrates that the combination of both graph and algebraic distance based considerations can lead to the best results.
\subsection{Running time}
\par The implementation of stationary iterative processes and their running time are well studied issues. These topics are beyond the scope of this paper; we refer the reader to two books in which one can find discussions about sequential and parallel matrix-vector multiplications and general relaxations \cite{Grama,heath}. Typical running time of an AMG-based framework for linear ordering problems on graphs can be found in \cite{safro2003,safro2004,safro2005}. The introduction of the algebraic disctance did not increase significantly those running time estimations.

\section{Multiscale distance definition and hierarchical organization}\label{multiscale}
As mentioned in the introduction, the algebraic distance defined
above is only a crude {\it local distance}, measuring meaningful
relative distances only between neighboring nodes while also
detecting which nodes should not be considered as close neighbors.
This fuzzy local distance, which can be calculated rapidly, is
all we need for coarsening. A similar distance is then
calculated at each coarser level, thus yielding a multiscale
definition of distances through the entire graph, where at large
distances one defines the distances only between (usually large)
aggregates of nodes, not between any individual pair of distant
nodes. Such multiscale distances are not only far less expensive
to calculate: we next list several reasons why, {\it in principle},
distances in a general large graph should be {\it defined} better
in such a multiscale fashion.
\begin{itemize}
    \item At large distances the detailed individual distances (the exact travelling time from each house in Baltimore to each house in Boston, say)
    are usually not of interest.
    \item The distance in a general graph is a fuzzy notion, whose
    definition is to a certain  extent arbitrary. The difference
    between the two distances of two neighboring nodes from a
    third, far one is much less than the difference between various,
    equally legitimate distance definitions, and also far less than the
    accuracy of the graph data (e.g., its edge weights) and far
    less than the accuracy of solving the equations that define these
    distances.
    \item The most important reason: At different scales different
    factors should in principle enter into the distance
    definition. In particular, at increasingly larger distances,
    intrinsic properties of increasingly larger aggregates should
    play a progressively more important role. For example, in image
    segmentation, while at the finest level the ``closeness"  of
    two neighboring pixels (i.e., their chance to belong to the same
    segment) can be defined by their color similarity, at larger
    scales the closeness of two neighboring patches should be
    defined in terms of the similarity in their {\it average}
    color (which is different from the direct color similarity of
    neighboring pixels along the boundary between the patches) and also in
    terms of similarity in various texture measure (color variances,
    shape moments of subaggregates, average orientation of fine embedded
    edges, etc.) and other aggregative properties \cite{segm}, \cite{Nature}. Another example: in the
    problem of identifying clusters in a large set of points in
    $R^d$, at the finest level the distance between data points
    can simply be their Euclidean distance, while at coarser
    levels the distance between two aggregates of points should
    also take into account similarity in terms of aggregative
    properties, such as density, orientation and  dimensionality     \cite{Dan1}.
    \item The multiscale definition of distance also brings much
    needed flexibility into the way distances at one level are
    converted into distances at coarser levels. For example, in a
    graph whose finest level consists of face images and their
    similarity scores, if at some coarse level node $A$ is the
    union of  two fine-level nodes $A_1$ and $A_2$, and node
    $B$ is the union of $B_1$ and $B_2$, then the coarse weight
    $w_{AB}$ of the edge $(A,B)$ can be defined either as some
    {\it average} of $w_{A_1B_1}$, $w_{A_1B_2}$, $w_{A_2B_1}$, and
    $w_{A_2B_2}$, or alternatively as the {\it maximum} (or $L_p$
    average with large $p$) of those four weights. The former
    choice (average)  is more suitable if one wants to cluster
     faces having a {\it similar pose}, while the latter
    choice (max or $L_p$) is more suitable if we need clusters of
    images each belonging to the {\it same person} (or, generally,
    when the clustering should be based on transitive similarity).
\end{itemize}

An ingenious rigorous definition of distances in a general graph,
introduced in \cite{diffusion-maps}, is called {\it diffusion distance}. Denoting
by $p(t,y|x)$ the probability of a random walk on the graph
starting at $x$ to reach $y$ after $t$ steps, the diffusion
distance between two nodes $x_i$ and $x_j$ is defined by
\begin{equation}\label{diffusion}
d(x_i,x_j,t)^2 = \sum_y w(y)[p(t,y|x_i)-p(t,y|x_j)]^2~~,
\end{equation}
with some suitable choice of the node weights $w$. This is, in fact,
a multiscale definition of distance, with the diffusion time $t$
serving as the scaling parameter. And indeed the definition is
used for hierarchical organizations of graphs (even though
large-scale distances are still defined in detail for any pair of
nodes). The calculation of our ``algebraic distance" can be viewed
as just a fast way to compute a crude approximation to diffusion
distances at some small $t$.

The essential practical point is that this crude and inexpensive ``algebraic
distance" is all one needs for solving graph problems by repeated
coarsening. The calculation of the {\it diffusion map} (the diffusion
distances at various scales $t$) for a large graph is, on the other hand, quite expensive,
requiring computing (possibly many) eigenvectors of the graph Laplacian.
The fast way to calculate them should involve using a multiscale algorithm
such as AMG (which is likely to work well in those cases
where hierarchical organizations of the graph is meaningful; the
AMG solver can, by the way, calculate {\it many} eigenvectors for
nearly the same work of calculating only one \cite{Dan2}). However,
instead of calculating the diffusion map and then use it for
organizing the graph, {\it the AMG structure can itself be used directly,
and more efficiently for any such organization.}

Indeed, as pointed out in \cite{amg}, the same coarsening procedures used
by the AMG solver can directly be used for efficient
hierarchical organizations (such as multiscale clustering) of a graph
(as in \cite{Dan1}) or for multiscale segmentation of an image (as in \cite{segm}, \cite{Nature}). As
exemplified in this article (and also in \cite{safro2004}, \cite{safro2005}), this kind
of procedures can also be used for many other types of graph
problems, in particular, it can also be used for detecting small hidden cliques in random graphs \cite{feige-ron-2010}.

Thus, for discrete graphs, and analogously also for related
continuum field problems, although the diffusion map is a useful theoretical concept, it is often not the most practical
tool. We believe this to be true for most if not all spectral
graph methods (using eigenvectors of the graph Laplacian): The
same AMG structure that would rapidly calculate the eigenvectors can be
better used to directly address the problem at hand. As
pointed out in the discussion of multiscale distances, this can yield not just
faster solutions but also, and more important, better
definitions and more tunable treatments for many practical
problems.

\section{Conclusions}
We have proposed a new measure that quantifies the ``closeness" between two nodes in a given graph. The calculation of the measure is linear in the number of edges in the graph and involves just a small number of relaxation sweeps. The calculated measure is all that is required for coarsening purposes. A similar notion of distance is then calculated and used at each coarser level. We demonstrate the use of this new measure for the minimum (1,2)-sum
linear ordering problem and for the minimum 2-partitioning problem. The improvement in the results shows that this measure indeed detects the most important couplings in the graph and helps in producing a better coarsening, while at the same time preventing nonlocal vertices from belonging to the same coarse aggregate.

\section*{Acknowledgments}
This work was partially funded by the CSCAPES institute, a DOE project, and in part by DOE Contract DE-AC02-06CH11357. We express our gratitude to Dr. C\'edric Chevalier for providing us with a modification of Scotch software and helping to design the minimum 2-partitioning problem experiments.
%\section*{Appendix: Experimental setup}
%\begin{itemize}
%\item test set
%\item parameters
%\end{itemize}
%+++++++++++++++++++++++++++++++++++++++++++++++++++++++++++++++++++++++++++++++++++++++++++++++++++
%+++++++++++++++++++++++++++++++++++++++++++++++++++++++++++++++++++++++++++++++++++++++++++++++++++
\bibliographystyle{plain}
\bibliography{algdist}
\vspace*{1cm}
\hspace*{1.5in}{\scriptsize\framebox{\parbox{2.4in}{
The submitted manuscript has been created in part by UChicago Argonne, LLC, Operator of Argonne National Laboratory (``Argonne'').  Argonne, a U.S. Department of Energy Office of Science laboratory, is operated under Contract No. DE-AC02-06CH11357.  The U.S. Government retains for itself, and others acting on its behalf, a paid-up nonexclusive, irrevocable worldwide license in said article to reproduce, prepare derivative works, distribute copies to the public, and perform publicly and display publicly, by or on behalf of the Government.
}}}

\end{document}